\documentclass[12pt]{article}

\ifx\pdfoutput\undefined
\usepackage[dvips,bookmarks]{hyperref}
\else
\usepackage{hyperref}
\fi
\hypersetup{colorlinks=false,bookmarksopen,bookmarksnumbered,citecolor=blue,
   pdfstartview=FitH}

\usepackage[dvips]{graphicx}
\usepackage{latexsym}

\usepackage{amssymb,amsfonts,amsmath}
\usepackage{graphicx} 
\usepackage{indentfirst}

 \usepackage{bbm}

\parskip=\medskipamount

\newcommand {\cN}{{\cal N}}

\newcommand {\cS}{{\cal S}}

\def\a{\alpha}

\def\c{\chi}
\def\d{\delta}

\def\G{\Gamma}

\def\s{\sigma}

\def\F{\Phi}

\def\rd{{\rm d}}

\newcommand{\ad}{{\dot{\alpha}}}                           

\newcommand{\pa}{\partial}                           
\newcommand{\hf}{\frac12}

\newcommand{\be}{\begin{equation}}
\newcommand{\ee}{\end{equation}}
\newcommand{\bea}{\begin{eqnarray}}
\newcommand{\eea}{\end{eqnarray}}
\newcommand{\non}{\nonumber}
\def\double #1{#1{\hbox{\kern-2pt $#1$}}}

\begin{document}
\begin{titlepage}
\begin{flushright}
August, 2010 \\
\end{flushright}
\vspace{5mm}

\begin{center}
{\Large \bf  Variant supercurrents and Noether procedure}
\end{center}

\begin{center}
{\bf
Sergei M. Kuzenko\footnote{kuzenko@cyllene.uwa.edu.au}
} \\
\vspace{5mm}

\footnotesize{
{\it School of Physics M013, The University of Western Australia\\
35 Stirling Highway, Crawley W.A. 6009, Australia}}  
~\\
\vspace{2mm}

\end{center}
\vspace{5mm}

\begin{abstract}
\baselineskip=14pt
Consistent supercurrent multiplets are naturally associated with linearized off-shell supergravity models.
In arXiv:1002.4932 we presented the hierarchy of such supercurrents which correspond 
to all the models for linearized 4D $\cN=1$ supergravity classified a few years ago.
Here we analyze the correspondence between the most general supercurrent given
in arXiv:1002.4932 and the one obtained eight years ago in hep-th/0110131
using the superfield Noether procedure. We apply the Noether procedure to 
the  general $\cN=1$ supersymmetric nonlinear sigma-model and show that 
it naturally leads to the so-called $\cS$-multiplet, revitalized in arXiv:1002.2228.
\end{abstract}
\vspace{1cm}

\vfill
\end{titlepage}

\newpage
\renewcommand{\thefootnote}{\arabic{footnote}}
\setcounter{footnote}{0}
Inspired by a recent work of Komargodski and Seiberg \cite{KS2}, 
we have presented in \cite{K-var} the hierarchy of supercurrent multiplets
which are associated with the models for linearized 4D $\cN=1$ supergravity classified several years ago 
in \cite{GKP}. The most general form of such a multiplet is as follows:
\bea
{\bar D}^{\ad}{J} _{\a \ad} = { \c}_\a  +{\rm i}\,\eta_\a +D_\a X~,
\qquad {\bar D}_\ad {\c}_\a  &=& {\bar D}_\ad \eta_\a= {\bar D}_\ad {X}=0~, \non \\ 
\qquad D^\a {\c}_\a - {\bar D}_\ad {\bar {\c}}^\ad
&=&D^\a {\eta}_\a - {\bar D}_\ad {\bar {\eta}}^\ad = 0~.
\label{VIII}
\eea
Here $J_{\a \ad} = {\bar J}_{\a \ad} $ denotes the supercurrent, while the chiral superfields 
$\c_\a$, $\eta_\a$ and $X$ consitute the so-called multiplet of anomalies. 
The conservation law (\ref{VIII}) incorporates six smaller supercurrent multiplets, of which 
three include $12+12$ operators (minimal supercurrents) and the rest describe $16+16$ components
(reducible multiplets).

Let us recall the structure of the minimal supercurrent multiplets.
The case 
\bea
\c_\a = \eta_\a =0
\eea
describes the famous Ferrara-Zumino multiplet \cite{FZ}. It corresponds to the old minimal formulation 
for $\cN=1$ supergravity \cite{old}. Another choice 
\bea
X= \eta_\a =0
\label{3}
\eea
corresponds to the new minimal supergravity \cite{new} (this supercurrent was 
studied in \cite{tensor}). The third choice 
\bea
X=\c_\a  =0
\eea
corresponds to the minimal $12+12$  supergravity formulation
which was proposed a few years ago in \cite{BGLP}. Unlike the old minimal and the new minimal 
theories, this formulation is known at the linearized level only.

Among the three reducible supercurrents with $16+16$ components
 \cite{K-var}, the most interesting multiplet\footnote{The other reducible supercurrents are obtained 
 by setting either $\c_\a =0$ or  $X=0$. They appear to be less interesting than the one defined by eq. (\ref{6}), 
because the corresponding supergravity formulations  are known at the linearized level only.}
is singled out by the condition:
\bea
\eta_\a=0~.
\label{6}
\eea
It corresponds to the model (36) in \cite{GKP} which can be shown to be  a linearized version 
of the so-called $16+16 $ supergravity \cite{GGMW,LLO} known to be reducible \cite{Siegel16}.
After a `death sentence' given to this multiplet in the late 1970s, it was recently resurrected by 
Komargodski and Seiberg \cite{KS2}. These authors postulated
the following supercurrent conservation law
\bea
 {\bar D}^\ad J_{\a \ad} =  \c_\a +D_\a X~,
 \qquad {\bar D}_\ad {\c}_\a  = {\bar D}_\ad {X}=0~,  
\qquad D^\a {\c}_\a - {\bar D}_\ad {\bar {\c}}^\ad =0
\label{S-multiplet}
 \eea
 and proved, using laborious component calculations, its consistency
in the sense that $J_{\a\ad}$ contains a conserved energy-momentum tensor 
and a conserved supersymmetry current. The consistency of (\ref{S-multiplet})  
is  a built-in property within the approach of \cite{K-var}.
As argued in \cite{KS2}, the importance of the $\cS$-multiplet (\ref{S-multiplet})  
is that it exists for all known rigid supersymmetric theories, unlike the Ferrara-Zumino multiplet.

With regard to the most general supercurrent multiplet including $20+20$ operators, 
eq. (\ref{VIII}), it corresponds to 
a (two-parameter) sum  of the three minimal ($12+12$) linearized supergravity models\footnote{The 
gravitational superfield $H_{\a \ad}$ must be  one and the same in all of these three models.} 
listed in \cite{GKP}.  As shown in \cite{GKP}, such a theory is related to a linearized version of
the non-minimal formulation for $\cN=1$ supergravity \cite{non-min}.\footnote{The complex 
linear compensator $\G$ of non-minimal supergravity, ${\bar D}^2 \G=0$, 
can be represented as $\G=\s +G +{\rm i}\, F$, where $\s$ is a chiral scalar, ${\bar D}_\ad \s=0$, 
while  $G$ and $F$ are real linear superfields, ${\bar D}^2 G = G-\bar G =0$, and the same for $F$.
The constrained superfields $\s$, $G$ and $F$  describe the compensators emerging
in the three minimal supergravity formulations.}

Let us  represent the constrained chiral spinors $\c_\a$  and $\eta_\a$ in (\ref{VIII}) 
as vector-multiplet  field strengths, 
\begin{subequations}
\bea
\c_\a &=& -\frac{1}{4} {\bar D}^2 D_\a V~, \qquad \bar V =V~,
\label{cs1} \\
\eta_\a &=& -\frac{1}{4} {\bar D}^2 D_\a U~, \qquad \bar U=U~,
\label{cs2}
\eea
\end{subequations}
associated with prepotentials $V$ and $U$. Then, eq. (\ref{VIII}) turns into
\bea
{\bar D}^{\ad}{J} _{\a \ad} =
 -\frac{1}{4} {\bar D}^2 D_\a (V +{\rm i}\,U) +D_\a { X}~, \qquad
{\bar V} - V= {\bar U} -U= {\bar D}_\ad { X}=0~.
\label{VIII-mod}
 \eea
This  coincides with the general supercurrent derived  eight years ago
by Magro, Sachs and Wolf  \cite{MSW} with the aid of a modification of the superfield Noether procedure 
elaborated in \cite{Osborn} (see also \cite{Shizuya}), provided  the operators $V$ and $U$
are globally well-defined scalar superfields.
However, for such operators $V$ and $U$, 
 the supercurrent (\ref{VIII-mod}) proves to be  equivalent to the Ferrara-Zumino one.
Indeed,  in this case we can introduce
\bea
J^{(\rm FZ)}_{\a \ad} := J_{\a \ad} +\frac{1}{6} [D_\a , {\bar D}_\ad ] V-\pa_{\a\ad} U
~, \qquad
X^{(\rm FZ)}:= { X} -\frac{1}{12}{\bar D}^2  (V + 3 {\rm i} \,U)~,
\label{improve}
\eea
where $J^{(\rm I)}_{\a \ad} $ and $X^{(\rm FZ)}$ obey the 
conservation equation 
\bea
{\bar D}^{\ad}J^{(\rm FZ)}_{\a \ad} = D_\a X^{(\rm FZ)}~,\qquad {\bar D}_\ad X^{(\rm FZ)} =0~.
\label{conservation-old}
\eea
The obvious implication of this consideration (which appears to be implicit 
in the analysis of \cite{MSW,Osborn}) 
is that any rigid supersymmetric theory is characterized by a well-defined
Ferrara-Zumino supercurrent.

On the other hand, it has recently been argued in \cite{KS2,KS} that there exist 
rigid supersymmetric field theories 
for which the Ferrara-Zumino multiplet is not well-defined. 
Such theories include ({\sl i}) models with a  Fayet-Iliopoulos term; and  
({\sl ii})  $\cN=1$ nonlinear sigma-models with a non-exact K\"ahler form. 
 Let us discuss the second example that appears to be most interesting.\footnote{For
 theories with a Fayet-Iliopoulos term, the appropriate supercurrent is 
given by eq. (\ref{3})  \cite{DT,K-FI}.}
 Consider the general $\cN=1$ supersymmetric 
nonlinear sigma-model \cite{Zumino}
\bea
S =  \int \rd^8 z\,
 K(\Phi^{I},
 {\bar \Phi}{}^{\bar{J}})  +  \Big\{ \int {\rm d}^6z \,  W(\F^I) +{\rm c.c.} \Big\}~. 
\label{nact4}
\eea
The $\cS$-multiplet, eq. (\ref{S-multiplet}), for this model was found in \cite{KS2}. It is
\bea 
J_{\a \ad} = ({\bar D}_\ad {\bar \F}^{\bar J}) (D_\a \F^I) K_{I\bar J}~, \qquad 
\c_\a =-\hf {\bar D}^2 D_\a K~, \qquad X =-2 W~,
\label{S-multiplet2}
\eea
and thus $V=2K(\F , \bar \F)$ and $U=0$.
The three  operators in (\ref{S-multiplet2}) are clearly well-defined, in particular 
they are invariant under arbitrary K\"ahler transformations. However, since the K\"ahler 
potential $K$  is defined only locally in the target space, the operator $V$ is not globally well-defined
in general. In particular, if the K\"aher two-form of the target space is not exact, there is no
way to define the operator $V$ globally.
As a result, the improvement transformation (\ref{improve}) leads to ill-defined operators, 
and thus one is not allowed to use it.

We are going to demonstrate  that the Noether procedure 
does not force both  $V$ and $U$  to be globally well-defined operators.
Let us apply a simple version of  the Noether construction to the sigma-model (\ref{nact4}).
This model is super-Poincar\'e invariant. 
To start with, we  replace the standard  super-Poincar\'e transformation of $\F^I$ 
by  a general local variation  of the form:
\bea
\d \F^I = -\frac{1}{4} {\bar D}^2 \Big(L^\a D_\a \F^I\Big) ~,
\eea
where the parameter $L_\a (z) $ is an arbitrary spinor superfield.
Such a variation can be recognized as a general coordinate transformation 
of a covariantly chiral scalar superfield in $\cN=1$ supergravity (see textbooks
\cite{GGRS,BK} for reviews). In the special case when 
$L_\a$ is chosen to correspond to a super-Poincar\'e transformation, 
the action does not change. If  $L_\a$ is arbitrary, 
the variation of the action can be shown to be 
\bea
\d S=  \hf \int \rd^8 z\,L^\a
\Big\{   {\bar D}^\ad J_{\a \ad} 
-\c_\a -D_\a X
\Big\}
+{\rm c.c.}~,
\label{var}
\eea
where $J_{\a \ad}$, $\c_\a$ and $X$ are defined in (\ref{S-multiplet2}).
If the equations of motion
\bea
\frac{1}{4}{\bar D}^2 K_I  = W_I ~
\eea
hold, the above variation vanishes, $\d S=0$. Since the parameter $L^\a$ 
in (\ref{var}) is arbitrary, we obtain 
the supercurrent conservation equation (\ref{S-multiplet}).

There are two important lessons that we can immediately learn from the above simple calculation.
First of all, the Noether supercurrent for the sigma-model (\ref{nact4}) coincides with the 
$\cS$-multiplet.
Secondly, the superfield Noether procedure is flexible enough in the sense that it
does not force both prepotentials 
$V$ and $U$  in (\ref{VIII-mod})  to be globally well-defined operators.

The above sigma-model calculation is a streamlined version of the one described in \cite{MSW}.
Unfortunately, instead of giving the explicit expression for the $\cS$-multiplet, 
the authors of \cite{MSW} presented only the Ferrara-Zumino supercurrent of the sigma-model.
Nevertheless, it is fair to say that the existence of  the $\cS$-multiplet as a consistent supercurrent 
multiplet is a by-product of the results obtained in \cite{MSW,Osborn}.
It is due to the insight of Komargodski and Seiberg \cite{KS2} that the physical significance 
of this multiplet has been uncovered.

Many rigid supersymmetric theories can be coupled to one of the existing off-shell versions
of $\cN=1$ supergravity. In such cases, the Noether procedure is not really necessary, 
for a consistent (rigid) supercurrent can be computed by varying the (curved-space) action 
with respect to the supergravity prepotentials, and then switching off the supergravity background.
However, there exist theories for which (i) no coupling to supergravity is possible; or 
(ii) such a coupling is not known. For example,  the sigma-model (\ref{nact4}) can not 
consistently  be coupled 
to supergravity unless the target space is Hodge-K\"ahler \cite{WB}.
On the other hand, it is not known how to couple supergravity to
the off-shell gauge models for higher-spin massless multiplets 
\cite{KSP} (see \cite{BK} for a review).
In all such cases, the Noether procedure \cite{MSW,Osborn} becomes indispensable.

It was conjectured in \cite{KS2} that the $\cS$-multiplet, eq.  (\ref{S-multiplet}), exists for all rigid 
supersymmetric theories. 
On the other hand, both the linearized supergravity analysis and the Noether procedure allow for 
the more general conservation equation (\ref{VIII}). It would be interesting to understand whether 
it is always possible, modulo an improvement transformation,  to set $\eta_\a=0$ or not.
\\

\noindent
{\bf Acknowledgements:}\\
The author is grateful to Ivo Sachs for reminding him of  \cite{MSW}. 
This work  is supported in part by the Australian Research Council.

\small{

}

\end{document}